\documentclass[ALICE,manyauthors]{cernphprep}

\usepackage[comma,square,numbers,sort&compress]{natbib}
\usepackage{hyperref}
\usepackage{lineno}
\usepackage{color}

\newcommand{\sNNp}{Pb--Pb collisions at $\sqrt{s_{\rm{NN}}} = 5.02$ TeV}

\newcommand{\sNNps}{Pb--Pb collisions at $\sqrt{s_{\rm{NN}}} = 5.02$ TeV }
\newcommand{\sNNxs}{Xe--Xe collisions at $\sqrt{s_{\rm{NN}}} = 5.44$ TeV }
\newcommand{\es}{$\eta/s$}

\begin{document}
\begin{titlepage}
\PHyear{2018}
\PHnumber{090}      
\PHdate{30 April}  

\title{Anisotropic flow in Xe--Xe collisions at $\mathbf{\sqrt{s_{\rm{NN}}} = 5.44}$ TeV}
\ShortTitle{Anisotropic flow in Xe--Xe collisions at $\sqrt{s_{\rm{NN}}} = 5.44$ TeV}   

\Collaboration{ALICE Collaboration\thanks{See Appendix~\ref{app:collab} for the list of collaboration members}}
\ShortAuthor{ALICE Collaboration} 

\begin{abstract}
The first measurements of anisotropic flow coefficients $v_{\rm{n}}$ for mid-rapidity charged particles in \sNNxs are presented. Comparing these measurements to those from \sNNp, $v_{2}$ is found to be suppressed for mid-central collisions at the same centrality, and enhanced for central collisions. The values of $v_{3}$ are generally larger in Xe--Xe  than in Pb--Pb at a given centrality. These observations are consistent with expectations from hydrodynamic predictions. When both $v_{2}$ and $v_{3}$ are divided by their corresponding eccentricities for a variety of initial state models, they generally scale with transverse density when comparing Xe--Xe and Pb--Pb, with some deviations observed in central Xe--Xe and Pb--Pb collisions. These results assist in placing strong constraints on both the initial state geometry and medium response for relativistic heavy-ion collisions.
\end{abstract}
\end{titlepage}
\setcounter{page}{2}

\section{Introduction}

Relativistic heavy-ion collisions are believed to create a Quark--Gluon Plasma (QGP), a state of matter consisting of deconfined color charges. The pressure gradients in the QGP medium convert spatial anisotropies in initial conditions of the collision to momentum anisotropies of produced particles via multiple interactions, a phenomenon referred to as {\it anisotropic flow} \cite{Ollitrault:1992bk}. The magnitude of anisotropic flow can be characterized by the flow coefficients ($v_{\rm{n}}$), which are obtained from a Fourier expansion of the angular distribution of produced particles \cite{Voloshin:1994mz}
\begin{equation} 
\label{eq:v_n}
\frac{\rm{d}\it{N}}{\rm{d}\varphi} \propto 1 + 2\sum_{\rm{n}=1}^{\infty} v_{\rm{n}} \cos [\rm{n}(\varphi-\Psi_{\rm{n}})],
\end{equation}  
where $\varphi$ is the azimuthal angle of the produced particle, n is the flow harmonic, and $\Psi_{\rm{n}}$ is the corresponding symmetry plane angle. For the second and third order flow coefficients ($v_2$ and $v_3$), various hydrodynamical calculations have demonstrated the approximate relation \cite{Holopainen:2010gz, Qin:2010pf, Qiu:2011iv, Gale:2012rq, Niemi:2012aj}
\begin{equation} 
v_{\rm{n}} \approx \kappa_{\rm{n}} \, \varepsilon_{\rm{n}},
\label{eq:vk}
\end{equation} 
where $\varepsilon_{\rm{n}}$ is the corresponding eccentricity coefficient, which governs the shape of the initial state. The variable $\kappa_{\rm{n}}$ encodes the response of the medium, and in particular is sensitive to the shear viscosity over entropy density ratio (\es) and the lifetime of the system. When values of $\eta/s$ are finite, this inhibits the development of momentum anisotropies. It has also received a broader interest, as its lower bound is different for perturbative QCD \cite{Huot:2006ys} and AdS/CFT \cite{Kovtun:2004de}. Experimental data from both the Relativistic Heavy-Ion Collider (RHIC) and the Large Hadron Collider (LHC) \cite{Back:2004je, Adams:2005dq, Adcox:2004mh, ALICE:2011ab, ATLAS:2012at, Chatrchyan:2013kba, Adam:2016izf}, have implied values of $\eta/s$ close to the AdS/CFT minimum of $1/4\pi$ \cite{Kovtun:2004de}, suggesting that the QGP behaves as a near perfect fluid. However, uncertainties in the modeling of the initial state have prevented the extraction of more precise information \cite{Song:2010mg, Heinz:2013th, Song:2017wtw}. 

The data set from the LHC Xe--Xe run completed in 2017 may provide an opportunity to further constrain $\eta/s$. For mid-central collisions, various initial state models predict \sNNxs and \sNNps have similar values of $\varepsilon_{2}$ at a given centrality \cite{Eskola:2017bup, Giacalone:2017dud}. However, at the same centrality the Xe--Xe system size is smaller than Pb--Pb, and the impact of a finite $\eta/s$ suppresses  $\kappa_{2}$ by $1/R$, where $R$ corresponds to the transverse size of the system \cite{Giacalone:2017dud}. Therefore, ratios of Xe--Xe/Pb--Pb $v_{2}$ coefficients in the mid-centrality range could be directly sensitive to $\eta/s$, with the influence of the initial state largely cancelling out. Furthermore, hydrodynamical calculations have shown that $v_{\rm{n}}/\varepsilon_{\rm{n}}$ increases monotonically with the transverse density $1/S$ d$N_{\rm{ch}}$/d$\eta$ (d$N_{\rm{ch}}$/d$\eta$ is the charged particle density and $S$ is the transverse area) across different collision energies and systems \cite{Voloshin:1999gs, Song:2008si, Song:2010mg}. Both $\varepsilon_{\rm{n}}$ and $S$ are quantities that are obtained from an initial state model. A violation of the scaling can be the result of incorrect modeling of the density ($S$) or the azimuthal geometry ($\varepsilon_{\rm{n}}$). That being the case, such an exercise where one compares $v_{\rm{n}}/\varepsilon_{\rm{n}}$ as a function of  $1/S$ d$N_{\rm{ch}}$/d$\eta$ for both Xe--Xe and Pb--Pb collisions can further constrain the initial state. Similar investigations using RHIC data from Cu--Cu and Au--Au collisions led to important refinements in this regard, such as the relevance of initial state fluctuations \cite{Alver:2006wh, Alver:2007qw, Alver:2010rt} and realization of finite values of $\varepsilon_{\rm{n}}$ for higher order odd values of n ($\rm{n}\geq3$) \cite{Alver:2010gr}. On the other hand, an observed violation of this scaling using experimental data (assuming the initial state predictions are accurate) may reveal deficiencies in the aforementioned hydrodynamical modeling. Addressing how the information from Xe--Xe collisions can shed more light on both the medium response and initial state, is the central goal of this Letter.

\section{Analysis details}
The two data sets analyzed were recorded by the ALICE detector at the LHC during the Xe--Xe (2017) and Pb--Pb (2015) runs at the center of mass energies of $\sqrt{s_{\rm NN}}=5.44$~TeV and $\sqrt{s_{\rm NN}}=5.02$~TeV, respectively. A more detailed description of the ALICE detector and its performance can be found elsewhere~\cite{Aamodt:2008zz, Abelev:2014ffa,Acharya:2018hhy}. Charged-particle tracks at mid-rapidity are reconstructed using the Time Projection Chamber (TPC)~\cite{Aamodt:2008zz, Alme:2010ke}, the primary tracking detector. Information from the Inner Tracking System (ITS)~\cite{Aamodt:2008zz, Aamodt:2010aa} is used to improve the spatial and momentum resolution of the TPC tracks. This helps to reject the background from secondaries, which originate from weak decays, conversions, secondary hadronic interactions in the detector material, and pile-up. The two innermost layers of the ITS, the Silicon Pixel Detector (SPD), are employed for triggering and event selection. The two V0 counters~\cite{Aamodt:2008zz, Abbas:2013taa}, each containing 32 scintillator tiles and covering $2.8<\eta<5.1$ (V0A) and $-3.7<\eta<-1.7$ (V0C), provide information for triggering, event selection, the determination of centrality and the symmetry plane angle ~\cite{Abelev:2013qoq}. The trigger conditions and the event selection criteria are described elsewhere \cite{Abelev:2014ffa}. An offline event selection is applied to remove beam-induced background (i.e.,\ beam-gas events) and pile-up events, which are rejected using information from the ITS and V0 detectors. Primary vertex information is provided by tracks reconstructed in the ITS and TPC. Only events with a reconstructed primary vertex within 10~cm from the center of the detector along the beam axis (that position is denoted by $PV_z$) are used in the analysis to ensure a uniform acceptance in $\eta$. The resulting event sample available for analysis consisted of $\sim 1.0$M Xe--Xe events in the 0--70\% centrality range, and $\sim 67$M events for Pb--Pb collisions in the same centrality interval. 

The charged tracks at mid-rapidity used to determine the flow coefficients have the kinematic values $0.2<p_{\rm{T}}<10$~GeV/$c$ and $|\eta|<0.8$. The track fit uses an SPD hit if one exists within the trajectory, if not, they are constrained to the primary vertex. Such a configuration leads to a relatively flat azimuthal acceptance. Track quality is ensured by requiring tracks to have at least 70 TPC space points out of a maximum of 159 with an average $\chi^2$ per degree-of-freedom for the track fit lower than 2. In addition, the distances of closest approach to the primary vertex in the $xy$ plane and $z$ direction are required to be less than 2.4 cm and 3.2 cm, respectively. The charged particle track reconstruction efficiency is estimated from HIJING simulations~\cite{Wang:1991hta, Gyulassy:1994ew} combined with a GEANT3~\cite{Brun:1994aa} transport model. 

In order to extract the flow coefficients from charged particles produced in either Xe--Xe or Pb--Pb collisions, the Scalar Product \cite{Adler:2002pu} and Generic Framework \cite{Bilandzic:2010jr, Bilandzic:2013kga} methods are used, which evaluate $m$ particle flow coefficients $v_{\rm n}\{m\}$. The $v_{\rm n}\{m\}$ coefficients characterize flow fluctuations, and are sensitive to correlations not related to the common symmetry planes $\Psi_{\rm n}$ (``non-flow''), such as those due to resonances and jets. The contribution from flow fluctuations was shown to decrease $v_{\rm n}\{m \geq 4\}$ and increase $v_{\rm n}\{2\}$ relative to  $\langle v_{\rm n} \rangle$ ~\cite{Voloshin:2008dg}. In the absence of flow fluctuations and non-flow, $v_{\rm n}\{m\}$ is independent of $m$. Both methods feature calculations involving the ${\bf Q}_{\rm n}$-vector which is defined as
\begin{equation}
\label{eq:Qn}
{ \bf Q}_{\rm{n}}=\sum_{i}^{M} e^{\rm{in}\varphi_{\it{i}}},
\end{equation}
where $M$ is the number of particles used to build the ${\bf Q}_{\rm n}$-vector in a single event, and $\varphi_{i}$ is the azimuthal angle of particle $i$. For the Scalar Product method, the flow coefficients $v_{\rm n}$ (denoted as $v_{\rm n}\{2, |\Delta\eta|>2\}$) are measured using
\begin{equation}
    v_{\rm{n}}\{{\rm SP}\} = \langle \langle {\bf u}_{{\rm{n}}, k} {\bf Q}_{\rm{n}}^{*} \rangle \rangle \Bigg/ \sqrt{ \frac{\langle {\bf Q}_{\rm{n}}  {\bf Q}_{\rm{n}}^{\rm A *} \rangle \langle  {\bf Q}_{\rm{n}} {\bf Q}_{\rm{n}}^{\rm B *} \rangle} { \langle  {\bf Q}_{\rm{n}}^{\rm A} {\bf Q}_{\rm{n}}^{\rm B *} \rangle } },
    \label{eq:mth_sp}
\end{equation}
where ${\bf u}_{{\rm{n}}, k} =\exp({\rm{in}}\varphi_k)$ is the unit flow vector of the particle of interest $k$. The brackets $\langle \cdots \rangle$ denote an average over all events, the double brackets $\langle \langle \cdots \rangle \rangle$ an average over all particles in all events, and $^*$ the complex conjugate. The vector ${\bf Q}_{\rm{n}}$ is calculated from the azimuthal distribution of the energy deposition measured in the V0A. Its $x$ and $y$ components are given by
\begin{equation}
 Q_{{\rm n},x} = \sum_j w_j \cos({\rm n} \varphi_j), \; Q_{{\rm n},y} = \sum_j w_j \sin({\rm n} \varphi_j),
\end{equation}
where the sum runs over all channels $j$ of the V0A detector ($j=1-32$), $\varphi_j$ is the azimuthal angle of channel $j$, and $w_j$ is the amplitude measured in channel $j$. The vectors ${\bf Q}_{\rm{n}}^{\rm A}$ and ${\bf Q}_{\rm{n}}^{\rm B}$ are determined from the azimuthal distribution of the energy deposition measured in the V0C and the azimuthal distribution of the tracks reconstructed in the ITS and TPC, respectively. The large gap in pseudo-rapidity ($|\Delta\eta|>2.0$) between the charged particles in the TPC used to determine $v_{\rm n}$ and those in the V0A greatly suppresses non-flow effects. The course $\varphi$ segmentation of the V0 leads to a deterioration of resolution for higher order flow coefficients ($ \rm{n} \ge 4$), and prevents their measurements.

The flow coefficients $v_{\rm{n}}\{m\}$ from two- and multi-particle cumulants can also be obtained using the Generic Framework. The calculations using the ${\bf Q}_{\rm n}$-vector are generally much more complex than those shown in Eq. \ref{eq:mth_sp}, and can be found elsewhere \cite{Bilandzic:2013kga}. This approach provides a capability for the necessary corrections of systematic biases from non-uniform detector acceptance and tracking inefficiencies, and it has been used in other measurements ~\cite{Adam:2016izf,ALICE:2016kpq,Acharya:2017gsw}.  It can also be used to suppress non-flow by placing an $\eta$-gap between various ${\bf Q}_{\rm n}$-vectors. The non-flow contribution to $v_{\rm n}\{m \geq 4\}$ in this framework is strongly suppressed by construction without the use of an $\eta$-gap. The newly developed sub-event methods \cite{Jia:2017hbm, Huo:2017nms}, provide additional means of suppressing any residual non-flow contributions for $v_{\rm n}\{m \geq 4\}$. The Generic Framework is used for measurements of $v_{\rm n}\{2, |\Delta\eta|>1\}$ and $v_{\rm n}\{m \geq 4\}$ (including $v_{2}\{4,\text{3 sub-event}\}$) from charged tracks in the TPC acceptance only.

When constructing Eq. \ref{eq:Qn} from charged particles to determine $v_{\rm{n}}\{m\}$, particle-wise weights are placed to account for non-uniformities in the $\varphi$ acceptance and $p_{\rm{T}}$ dependent efficiencies. The systematic uncertainties for $v_{\rm{n}}\{m\}$ have three sources: event selection, track type/selection, and the ${ \bf Q}_{\rm{n}}$-vector correction procedure. The event selection contributions were determined by varying the $PV_{z}$ ranges, not applying the pile-up rejection criteria, and using a different detector system (ITS) for centrality determination. The track type/selection uncertainties were determined by using tracks with TPC information only or tracks that always have an ITS hit (which changes the contributions from secondary particles), changing the track quality cuts (such as the minimum number of TPC space points), and comparing any differences between determining ${ \bf Q}_{\rm{n}}$ or ${\bf u}_{n, k}$ from positive or negative only TPC tracks (both charge signs are used to build a flow vector for the final results). Finally, the uncertainties in ${ \bf Q}_{\rm{n}}$-vector correction procedure contribution are due to uncertainties in the $p_{\rm{T}}$ dependent efficiencies. The individual sources of systematic uncertainty are assumed uncorrelated and are added in quadrature to obtain the overall estimated systematic uncertainties. For the $p_{\rm{T}}$-integrated $v_{\rm{n}}\{m\}$ coefficients, the total systematic uncertainties are typically 2--3\%, and smaller than the marker size in the corresponding figures. The systematic uncertainties for the $p_{\rm{T}}$-differential coefficients can be larger, and are denoted by boxes in the relevant figures. 

\begin{figure}[h]
\begin{center}
\includegraphics[width = 0.6\textwidth]{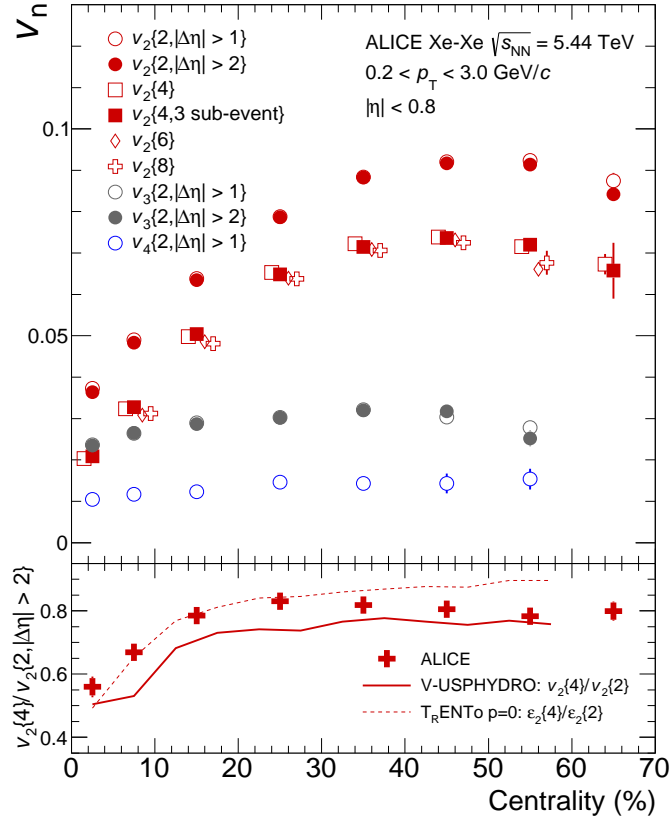}
\caption{Top panel: Charged particle $v_{\rm n}$ integrated over the transverse momentum range $0.2<p_{\rm{T}} <3.0$~GeV/$c$ as a function of centrality from Xe--Xe collisions. The various techniques are explained in the text. Only statistical uncertainties are visible (thin vertical lines). Bottom panel: Ratios of $v_{2}\{4\}$/$v_{2}\{2\}$ compared to some theoretical predictions. The hydrodynamic predictions use a shear viscosity over entropy ratio $\eta/s=0.047$ and  initial conditions from the T$_{\rm{R}}$ENTo model \cite{Giacalone:2017dud,Moreland:2014oya}. For $v_{2}\{2\}$, the ALICE measurements implement a $|\Delta\eta|>2.0$ gap which is not used in the models. }
\label{Fig1}
\end{center}
\end{figure}

\section{Results}

The top panel of Fig.\  \ref{Fig1} shows two- and multi-particle $p_{\rm{T}}$-integrated $v_{\rm{n}}\{m\}$ coefficients from \sNNxs as a function of centrality. A stronger dependence of $v_{2}$ with centrality compared with $v_{3}$ or $v_{4}$, also observed in Pb--Pb collisions at LHC energies \cite{ALICE:2011ab, ATLAS:2012at, Chatrchyan:2013kba, Adam:2016izf}, is expected based on simple considerations of how the almond shaped overlap region changes with centrality for A--A collisions. Given that near-side non-flow correlations (where the particles involved have similar values of $\varphi$ and $\eta$) are expected to be the largest non-flow contribution, the similarities observed for $v_{\rm{n}}\{2,|\Delta \eta| > 2\}$ and $v_{\rm{n}}\{2,|\Delta \eta| > 1\}$ indicate non-flow is strongly suppressed by a gap of one unit of pseudorapidity. The extracted values of $v_{2}\{m\geq 4\}$ use ${\bf Q}_{\rm{n}}$-vectors without any $\eta$ gaps. The $v_{2}\{4,\text{3 sub-event}\}$ results have $\eta$ gaps between the ${\bf Q}_{\rm{n}}$-vectors to suppress non-flow. The sub-event regions are $-0.8 < \eta \leq -0.4$, $-0.4 < \eta \leq 0.4$ and $0.4 < \eta \leq 0.8$. The equivalence with $v_{2}\{4\}$ (no $\eta$ separation) demonstrates that such a gap is actually not required for these flow coefficients. Given all those observations regarding non-flow, one can interpret differences between $v_{2}\{2\}$ and $v_{2}\{4\}$, $v_{2}\{6\}$, $v_{2}\{8\}$ to be largely driven by flow fluctuations \cite{Voloshin:2008dg}.
To quantify these differences, in the bottom panel of Fig.\  \ref{Fig1}, the ratio $v_{2}\{4\}$/$v_{2}\{2,|\Delta \eta| > 2\}$ is shown, which is found to decrease for central collisions. The results are compared to a hydrodynamic calculation in the same panel, which uses an $\eta/s=0.047$ to model the medium response \cite{Giacalone:2017dud}. For these hydrodynamic calculations, the T$_{\rm{R}}$ENTo initial condition model \cite{Moreland:2014oya} is used to determine the eccentricities. The justification for using $\rm{p}=0$ is described later in the Letter. The initial condition model implements a $^{129}$Xe $\beta_{2}$ deformation ($\beta_{2}=0.162$), which is predicted for the $^{129}$Xe nucleus \cite{Moller:2015fba}, but has never been measured directly. It modifies the Woods-Saxon distribution as follows \cite{Hagino:2006fj}

\begin{equation}
\label{eq:b2}
\rho(r,\theta) = \frac{\rho_{0}}{1+e^{(r-R_{0}-R_{0}\beta_{2}Y_{20}(\theta))/a}},
\end{equation}
where $\rho_{0}$ is the density at the center of the nucleus, $R_{0}$ the nuclear radius, $r$ is the distance away from the center, $Y_{20}$ is a Bessel function of the second kind, and $a$ is the skin depth. According to Eq. $\ref{eq:vk}$, the ratio of flow coefficients $v_{2}\{4\}$/$v_{2}\{2\}$ should be identical to the ratio of initial state eccentricities $\varepsilon_{2}\{4\}/\varepsilon_{2}\{2\}$. To test this relation, the bottom panel of Fig.\  \ref{Fig1} also shows the flow coefficient ratios and the eccentricity ratios from the same model. The difference between the two curves shows that Eq. $\ref{eq:vk}$ only holds approximately. The hydrodynamic calculations generally predict lower ratios compared to the data, with the largest deviations being in the semi-central region (10--50\%).

\begin{figure}[t]
\begin{center}
\includegraphics[width = 0.6\textwidth]{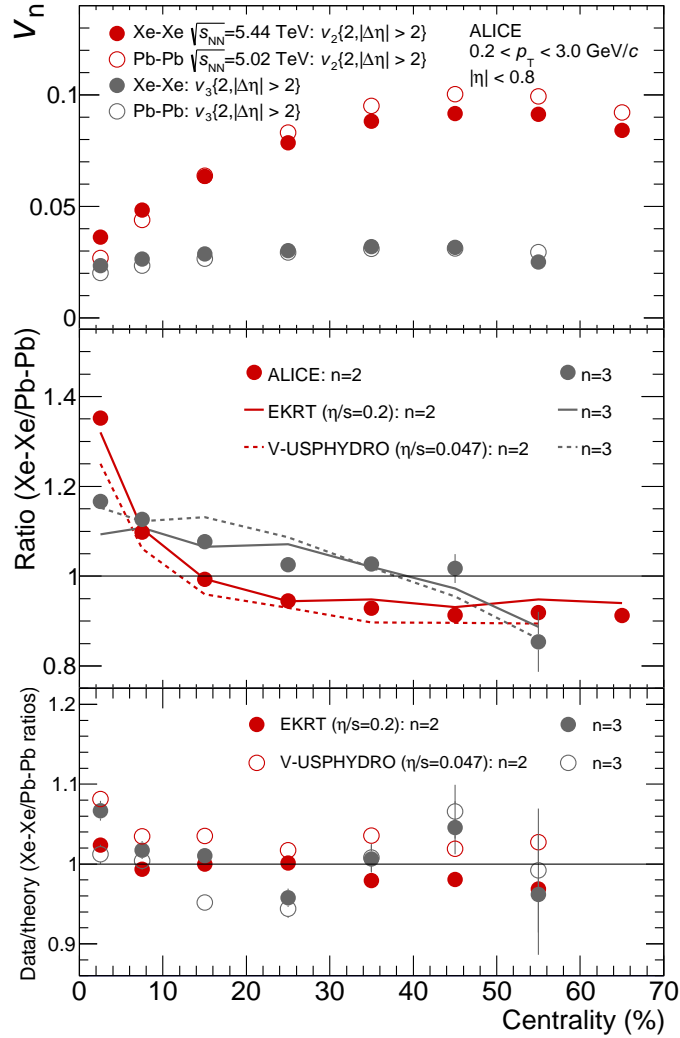}
\end{center}
\caption{Top panel: Comparisons of charged particle $v_{\rm n}\{2\}$ integrated over the transverse momentum range $0.2<p_{\rm{T}} <3.0$~GeV/$c$ as a function of centrality from Xe--Xe and Pb--Pb collisions. Middle panel: Ratio of $v_{\rm n}\{2\}$(Xe--Xe/Pb--Pb) coefficients. Bottom panel: Double ratio of data and theory. Hydrodynamical model predictions from EKRT \cite{Eskola:2017bup} and V-USPHYDRO \cite{Giacalone:2017dud} are shown. In all cases, only statistical uncertainties are visible (thin vertical lines).}
\label{Fig2}
\end{figure}

\begin{figure}[t]
\begin{center}
\includegraphics[width = 0.6\textwidth]{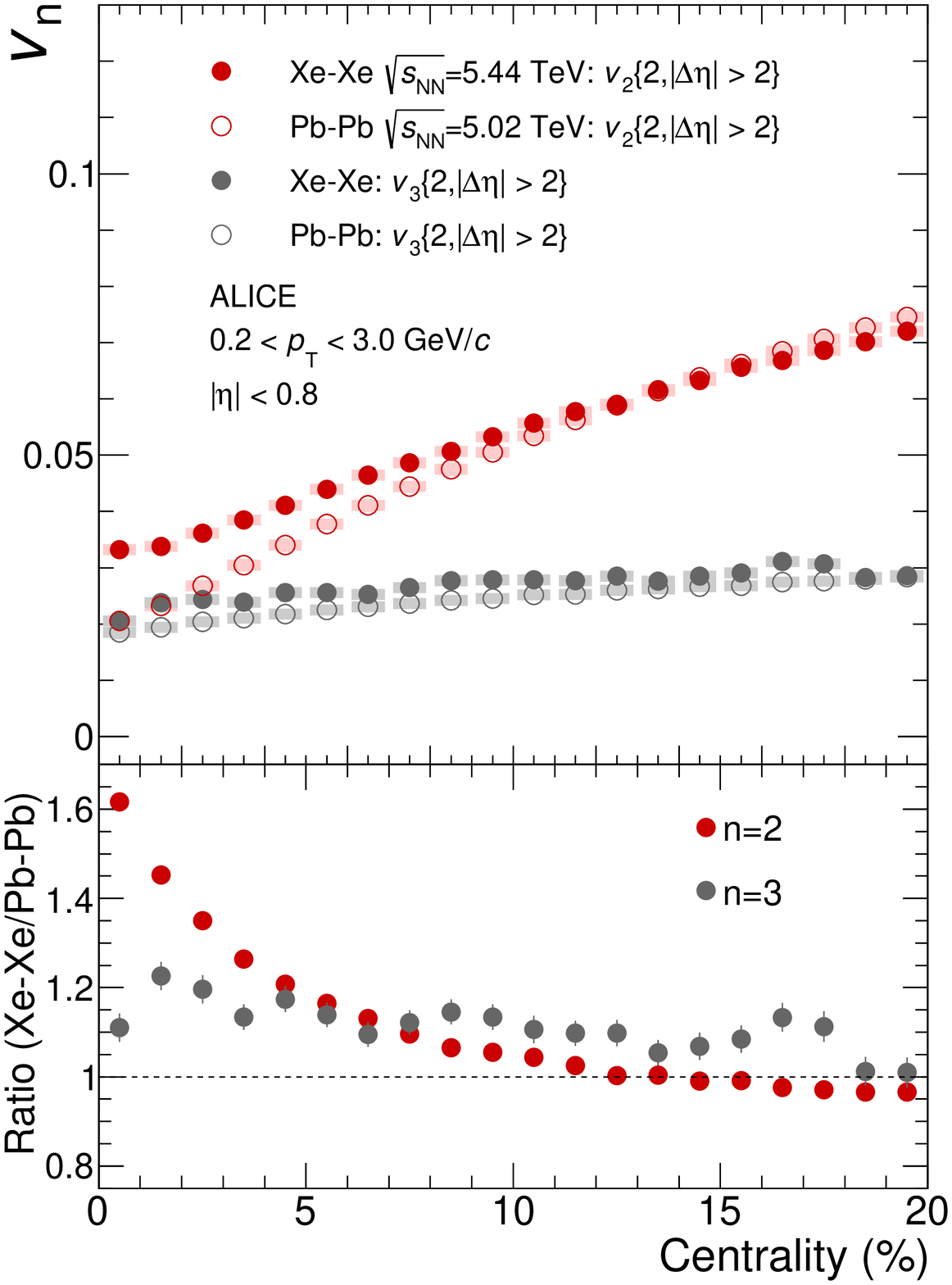}
\end{center}
\caption{Top panel: Comparisons of charged particle $v_{\rm n}\{2\}$ integrated over the transverse momentum range $0.2<p_{\rm{T}} <3.0$~GeV/$c$  from Xe--Xe and Pb--Pb collisions for finer centrality bins in central collisions. Statistical and systematic uncertainties are shown as lines and boxes, respectively. Bottom panel: Corresponding ratio of $v_{\rm n}\{2\}$(Xe--Xe/Pb--Pb) coefficients.}
\label{Fig2e}
\end{figure}

Figure \ref{Fig2} shows comparisons of two-particle $p_{\rm{T}}$-integrated $v_{\rm{n}}\{2\}$ coefficients from Xe--Xe and Pb--Pb collisions as a function of centrality. The differences between the two systems are typically within 10\% except for $v_{2}\{2\}$ in central 0--5\% collisions where the Xe--Xe values are $\sim$ 35\% higher. For the V-USPHYDRO and EKRT models \cite{Eskola:2017bup, Giacalone:2017dud} shown, both sets of the used initial condition models demonstrate $\varepsilon_{2}\{2\}$(Xe--Xe)/$\varepsilon_{2}\{2\}$(Pb--Pb) $\sim$ 1 for the semi-central range 20--60\% (not shown in the figure). However, $v_{2}\{2\}$(Xe--Xe)/$v_{2}\{2\}$(Pb--Pb) $\sim$ 0.9 from the data, which might be the result of the viscous effects described in the Introduction. When implementing the hydrodynamical response, both models also show a similar suppression for the smaller Xe--Xe system, albeit with differences of up to $\sim$ 5\% compared to the data. On the other hand, despite using different values of $\eta/s$, both models predict similar ratios in the semi-central range. Some assumptions used in each of the models (such as the freeze-out temperature) are different, and investigating the impact of those assumptions on $v_{2}\{2\}$(Xe--Xe)/$v_{2}\{2\}$(Pb--Pb) ratio should be a topic of further theoretical investigations. 

Both sets of model predictions (V-USPHYDRO and EKRT) implement a $^{129}$Xe deformation using $\beta_{2}=0.162$. The value of $\beta_{2}$ is zero for the $^{208}$Pb nucleus, as it is a double magic nucleus. The deformation for the Xe--Xe V-USPHYDRO predictions contributes $\sim$ 20\% to the observed $v_{2}\{2\}$ for central collisions (compared with the case where no deformation is implemented), and has no impact on $v_{2}\{2\}$ for centralities above 15\%. Regarding $v_{3}\{2\}$, it is generally larger in Xe--Xe, which reflects the fact that the initial conditions from both models show $\varepsilon_{3}\{2\}$(Xe--Xe) $> $ $\varepsilon_{3}\{2\}$(Pb--Pb) at a given centrality for the entire centrality range presented. The hydrodynamic predictions for $v_{3}\{2\}$ are similar for the two models, with maximum deviations of $\sim5$ \% from the data. The $\beta_{3}$ deformation for both the Xe and Pb nuclei is zero \cite{Moller:2015fba}, with both models assuming such a value.

In Fig.\  \ref{Fig2e}, similar comparisons are made in finer centrality bins as compared with Fig.\  \ref{Fig2} for central collisions. The transition where Xe--Xe  $v_{\rm{2}}\{2\}$ becomes larger than the Pb--Pb values occurs for a centrality of $\sim$ 15\%. For 0--1\% central collisions, where the overlap geometry is expected to play a minimal role for both systems, $v_{\rm{2}}\{2\}$ is $\sim$ 60\% larger for Xe--Xe collisions. In terms of the initial state, this is expected for two reasons. The first relates to the fact that the $^{129}$Xe nucleus is deformed while the $^{208}$Pb nucleus is not, and the second relates to the role of initial state fluctuations and the number of sources that contribute to $\varepsilon_{n}\{2\}$. It has been previously shown that $\varepsilon_{n}\{2\}$ decreases as the number of sources increases for a spherical system \cite{Bzdak:2013rya}, and if the number of sources were infinite, then $\varepsilon_{n}\{2\}$ would be zero in this centrality range. Given that a very central Pb--Pb collision is expected to have more sources than a very central Xe--Xe collision, fluctuations would be expected to give rise to larger values of $\varepsilon_{2}\{2\}$ for the latter. The same line of reasoning can explain why $v_{\rm{3}}\{2\}$ is observed to be larger in Xe--Xe compared to Pb--Pb in the same centrality interval.

\begin{figure*}[t]
\begin{center}
\includegraphics[width = 1\textwidth]{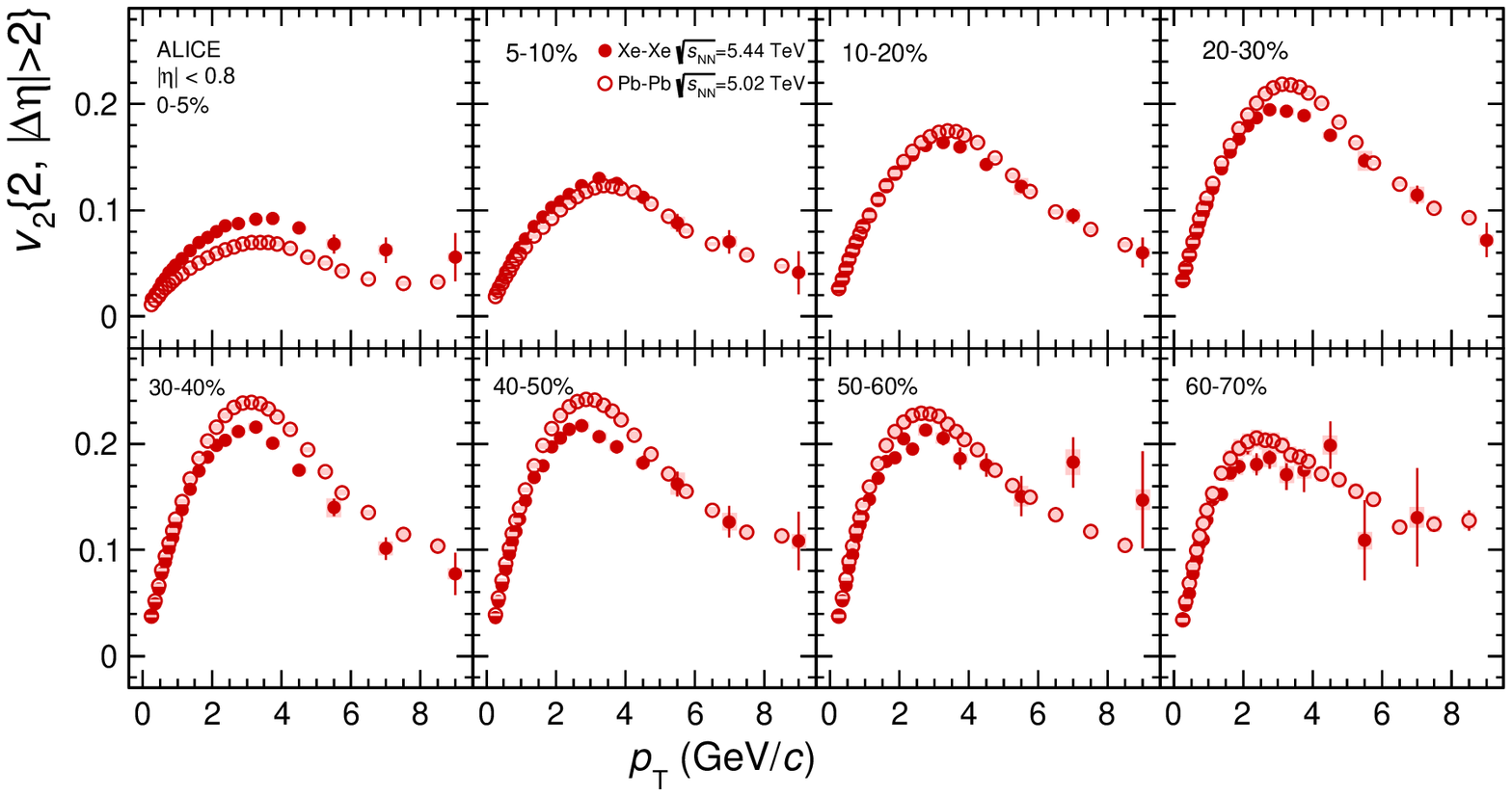}
\caption{The $p_{\rm{T}}$-differential $v_{2}$ for charged particles from \sNNxs and \sNNps for various centrality classes. Statistical and systematic uncertainties are shown as lines and boxes, respectively.}
\label{Fig3v2}
\end{center}
\end{figure*}
\begin{figure*}[t]
\begin{center}
\includegraphics[width = 1\textwidth]{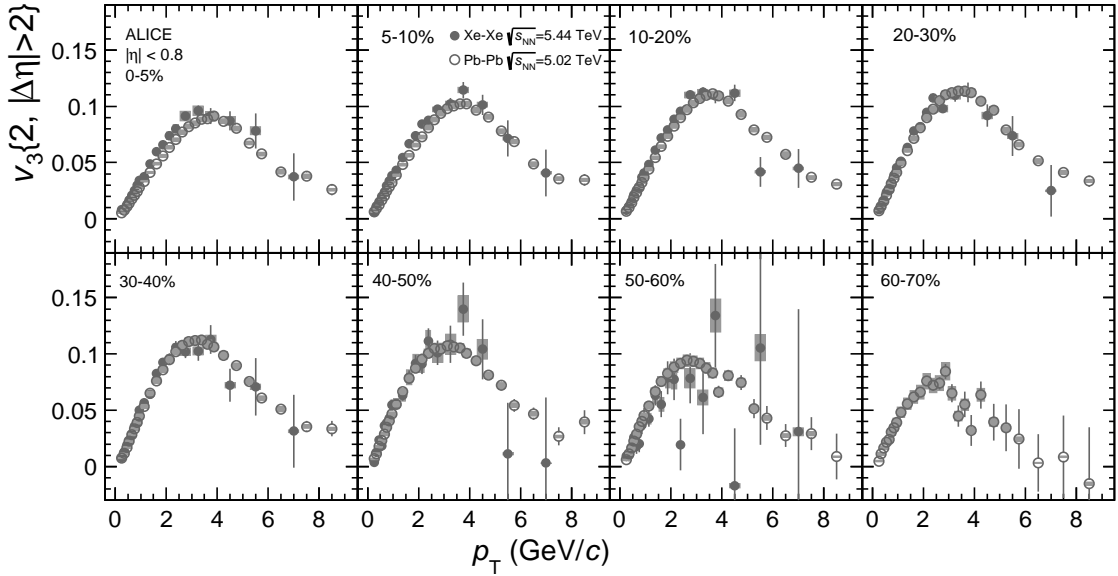}
\caption{The $p_{\rm{T}}$-differential $v_{3}$ for charged particles from \sNNxs and \sNNps for various centrality classes. Statistical and systematic uncertainties are shown as lines and boxes, respectively.}
\label{Fig3v3}
\end{center}
\end{figure*}

Figure \ref{Fig3v2} shows comparisons of two-particle $p_{\rm{T}}$-differential $v_{\rm{2}}\{2,|\Delta \eta| > 2\}$ coefficients from Xe--Xe and Pb--Pb collisions in various centrality bins. As mentioned, the larger $|\Delta \eta|$ gap measurements use both the TPC and the V0 detectors, which maximizes the number of particles used to build the ${\bf Q}_{\rm{n}}$-vectors. The corresponding reduction in statistical uncertainties is particularly useful for the higher $p_{\rm{T}}$ measurements. As expected, the centrality dependence of $v_{2}\{2,|\Delta \eta| > 2\}$ from Xe--Xe collisions follows that observed in Fig.\  \ref{Fig1}. Compared with Pb--Pb collisions in the semi-central bins, it appears the differences observed in Fig.\  \ref{Fig2} are larger in the mid-$p_{\rm{T}}$ region, and this will be investigated more quantitatively. Figure \ref{Fig3v3} shows the same comparison for $p_{\rm{T}}$-differential $v_{\rm{3}}\{2,|\Delta \eta| > 2\}$ coefficients. The Xe--Xe coefficients are typically larger than from Pb--Pb collisions at a given centrality at low $p_{\rm{T}}$, whereas the larger statistical uncertainties for the Xe--Xe coefficients at higher $p_{\rm{T}}$ make it difficult to establish whether there are any differences between the two systems.

\begin{figure}[t]
\begin{center}
\includegraphics[width = 1\textwidth]{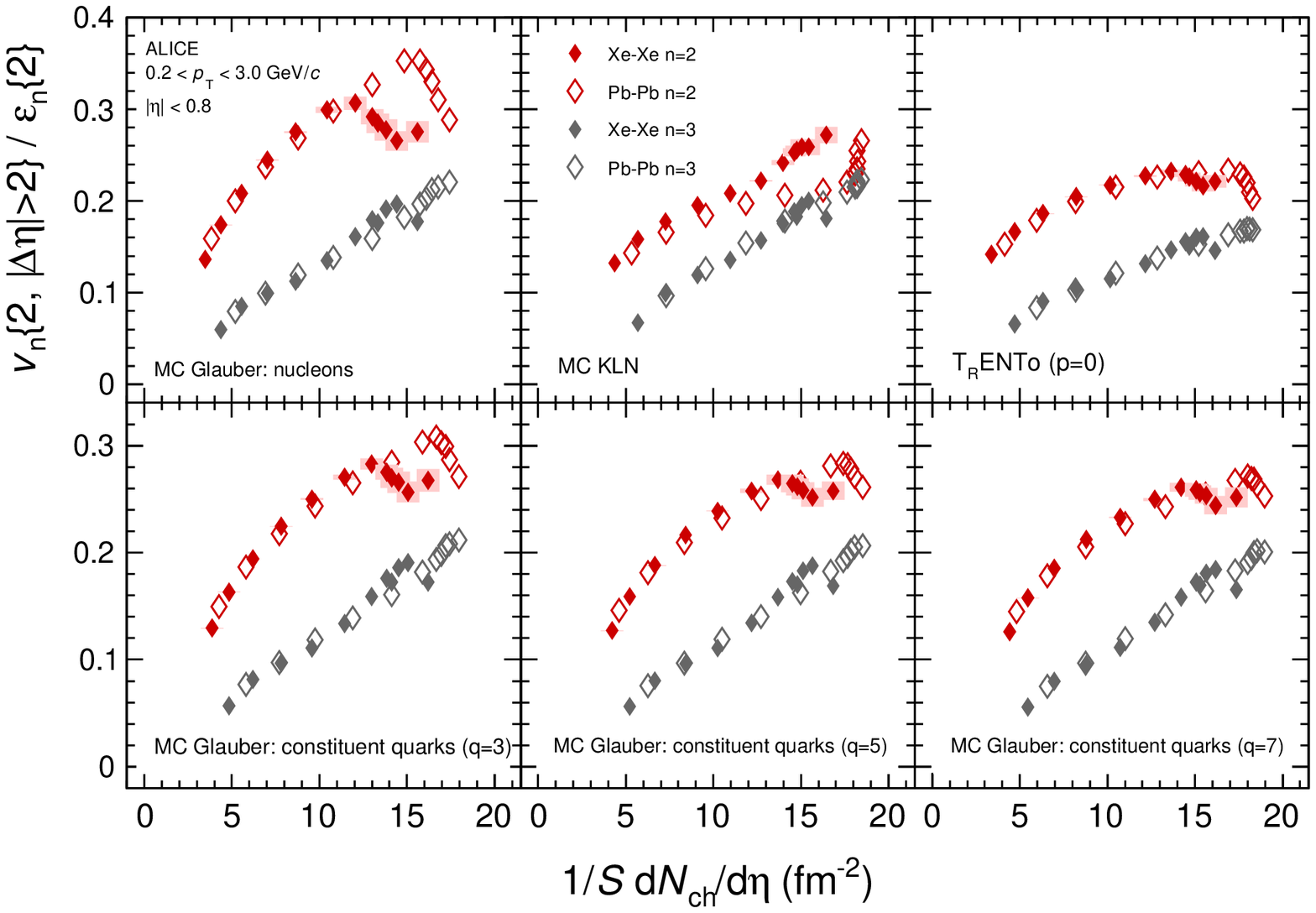}
\end{center}
\caption{Comparisons of $v_{\rm{n}}\{2\}/\varepsilon_{\rm{n}}\{2\}$ integrated over the transverse momentum range $0.2<p_{\rm{T}} <3.0$~GeV/$c$ as a function of  $1/S$ d$N_{\rm{ch}}$/d$\eta$ in Xe--Xe and Pb--Pb collisions, where $S$ and $\varepsilon_{\rm{n}}\{2\}$ are from various initial state models \cite{Moreland:2014oya, Loizides:2016djv, Drescher:2007ax}. The models are explained in the text. The $^{129}$Xe deformation implemented is $\beta_{2}=0.18\pm0.02$, with the box errors representing the uncertainty in $\beta_{2}$. Measurements of d$N_{\rm{ch}}$/d$\eta$ ($|\eta|< 0.5$) from Xe--Xe and Pb--Pb collisions were obtained from separate studies \cite{Acharya:2018hhy,Adam:2015ptt}.}
\label{Fig4}
\end{figure}

Figure \ref{Fig4} shows the $p_{\rm{T}}$-integrated $v_{\rm{n}}\{2\}/\varepsilon_{\rm{n}}\{2\}$ ratios as a function of $1/S$ d$N_{\rm{ch}}$/d$\eta$ in Xe--Xe and Pb--Pb collisions, where $S$ and $\varepsilon_{\rm{n}}\{2\}$ are obtained using various initial state models. The $v_{\rm{n}}\{2\}/\varepsilon_{\rm{n}}\{2\}$ ratio provides estimates of $\kappa_{\rm{n}}$ as per Eq. \ref{eq:vk}. As mentioned, when comparing $v_{\rm{n}}/\varepsilon_{\rm{n}}$ from different systems, a violation of the scaling with $1/S$ d$N_{\rm{ch}}$/d$\eta$ (which increases with centrality), maybe indicative of shortcomings in the modeling of the initial state (and its fluctuations). Regarding the model parameters used for this exercise, in the transverse plane for a single event, both the eccentricities and areas are calculated in the center of mass frame respectively according to

\begin{align}
\varepsilon_{\rm{n}} &= \frac{\sqrt{\langle r'^{\rm{n}} \cos(\rm{n}\phi')\rangle^2+\langle r'^{\rm{n}} \sin(\rm{n}\phi')\rangle^2}}{\langle r'^{\rm{n}} \rangle}, \\
S &= 4\pi\sigma_{x'}\sigma_{y'} ,
\label{eq:ac}
\end{align}
which is defined such that the sources that contribute to the eccentricity and area have the property $\langle x' \rangle = \langle y' \rangle = 0$, where $x',y'$ and $\varphi', r'$ are the cartesian and the polar coordinates of the source, respectively. The quantities $\sigma_{x'}$ and $\sigma_{y'}$  represent the standard deviations of the source distributions. The event averages used for Fig.\  \ref{Fig4} are $\varepsilon_{\rm{n}}\{2\}=\sqrt{\langle \varepsilon_{\rm{n}}\rangle^{2}+\sigma_{\varepsilon_{\rm{n}}}^{2}}$ and $\langle S \rangle$. The normalization of the area is chosen such that for a Gaussian distribution the average density coincides with $N_{part}/S$ ($N_{part}$ is the number of participating nucleons), and was used in a previous ALICE publication \cite{Adam:2015eta}. A deformation of $\beta_{2} = 0.18\pm0.02$ for the $^{129}$Xe nucleus is used \cite{Acharya:2018hhy, ALICE-PUBLIC-2018-003}. The value was obtained from extrapolating measurements of $\beta_2$ from nearby isotopes ($^{128}$Xe and $^{130}$Xe), and theoretical calculations \cite{Raman:2001fc, Zoltan:2015sc, Moller:2015fba}, with the uncertainty reflecting the different values obtained from each approach. The box errors in the figure represent the corresponding uncertainties on the ratio. For the Monte Carlo (MC) Glauber and KLN models, the values of $\varepsilon_{\rm{n}}\{2\}$ and $S$ for a given V0 based centrality class were extracted using a method described in a previous publication \cite{Abelev:2013qoq}. The multiplicity of charged particles in the acceptance of the V0 detector is generated according to a negative binomial distribution, based on the number of participant nucleons and binary collisions from each initial state model. The parameters used for this approach can be found elsewhere \cite{Adam:2015ptt, ALICE-PUBLIC-2018-003}, and were optimized to describe the multiplicity distribution from the data. Regarding the T$_{\rm{R}}$ENTo model, following other approaches \cite{Giacalone:2017dud, Moreland:2014oya}, the multiplicity in the acceptance of the V0 detector was modeled by scaling the entropy production, again to match the multiplicity distribution from the data.

The top left panel shows an investigation of such a scaling with the MC Glauber model \cite{Alver:2008aq, Loizides:2014vua}, which uses nucleon positions as the sources. In particular, for $v_{2}\{2\}$ in central Xe--Xe and Pb--Pb collisions, this model does not provide a clear scaling, and was already observed for $v_2$ from Au--Au and U--U collisions at RHIC using the same model \cite{Adamczyk:2015obl}. The scaling using the MC KLN model (version 32) \cite{Drescher:2007ax, ALbacete:2010ad}, which assumes gluon sources and uses a Color Glass Condensate approach to determine the gluon spatial distribution, is shown in the top middle panel. The MC KLN scaling appears to work well for $v_{3}\{2\}$, but fails for $v_{2}\{2\}$ with the Xe--Xe points being above Pb--Pb for more central collisions. A sudden rise is also observed for central Pb--Pb collisions. This behavior is in contrast to the MC Glauber nucleon model, where the Xe--Xe points are below Pb--Pb for central collisions. The top right panel investigates the scaling using the T$_{\rm{R}}$ENTo initial state model \cite{Moreland:2014oya}. In this model, the distribution of nuclear matter within the collision zone of A--A collisions is controlled by the p parameter, with $\rm{p}=0$ mimicking IP-Glasma initial conditions \cite{Schenke:2012wb, Schenke:2012fw}. The choice of parameter was determined using Bayesian statistics from a simultaneous fit of charged hadron multiplicity distributions, mean transverse momentum measurements, and integrated flow coefficients $v_{\rm n}$ in Pb--Pb collisions at $\sqrt{s_{\rm{NN}}} = 2.76 $ TeV \cite{Bernhard:2016tnd}. The IP-Glasma approach uses Color Glass Condensate calculations to determine the distribution of gluons in the initial state. This model provides a better scaling compared with the previous two other models. However for central Pb--Pb collisions, a drop is observed for $v_{2}\{2\}/\varepsilon_{\rm{2}}\{2\}$. The drop is also observed in the MC Glauber nucleon model, and appears to be present for the central Xe--Xe data. Such a drop is unexpected from hydrodynamic calculations \cite{Song:2008si, Song:2010mg}, which show a continuous increase of $v_{\rm{2}}/\varepsilon_{\rm{2}}$ with $1/S$ d$N_{\rm{ch}}$/d$\eta$. It may point to deficiencies in the initial state modeling of the regions in Xe--Xe and Pb--Pb collisions where initial state fluctuations play the largest role in generating second order eccentricities.

The bottom panels show ratios derived from constituent quark MC Glauber calculations, which use quarks contained in nucleons as the sources which contribute to the eccentricity \cite{Loizides:2016djv}. The parameter q refers to the number of constituent quarks per nucleon. All implementations of quark sources (3, 5, or 7) appear to give a reasonable scaling for $v_{2}$ and $v_{3}$, however some deviations are observed in central Xe--Xe and Pb--Pb collisions. The value $\rm{q}=5$ was found to describe the charged particle yields better than $\rm{q}=3$ at LHC energies (assuming the yields should scale with the total number of quarks) \cite{Loizides:2016djv}, and there are hints of a slightly better scaling with $\rm{q}=5$ for $v_2\{2\}$ in central Xe--Xe collisions compared to $\rm{q}=3$. These model implementations again show a drop for central Pb--Pb collisions, which is least pronounced for $\rm{q}=7$. This suggests initial state models need a higher number of sources per nucleon in order to achieve a continuous increase of $v_{\rm{2}}\{2\}/\varepsilon_{\rm{2}}\{2\}$ for more central Pb--Pb collisions, and a transverse density scaling when comparing Xe--Xe to Pb--Pb.

\begin{figure}[t]
\begin{center}
\includegraphics[width = 0.6\textwidth]{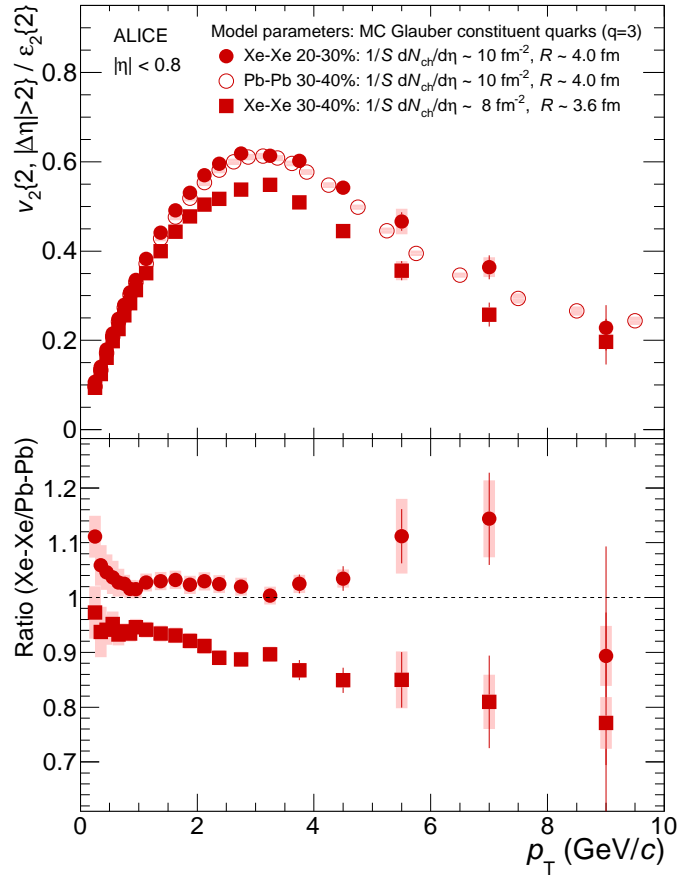}
\end{center}
\caption{Top panel: Comparison of $p_{\rm{T}}$-differential $v_{\rm{2}}\{2\}/\varepsilon_{\rm{2}}\{2\}$ from Xe--Xe and Pb--Pb collisions for a selection of centrality bins. Statistical and systematic uncertainties are shown as lines and boxes, respectively. Bottom panel: Ratios of the scaled coefficients from the top panel. The Pb--Pb points are interpolated in order to determine the ratio. The circle markers show Xe--Xe 20--30\%/Pb--Pb 30--40\% while the square makers show Xe--Xe 30--40\%/Pb--Pb 30--40\%.}
\label{Fig5}
\end{figure}
Finally, in Fig.\  \ref{Fig5}, an investigation of whether the transverse density scaling holds as a function of $p_{\rm{T}}$ is shown. Two Xe--Xe and Pb--Pb centrality bins with similar transverse densities ($1/S$ d$N_{\rm{ch}}$/d$\eta \sim 10$ fm$^{-2}$) are selected, and the $p_{\rm{T}}$-differential values of $v_{\rm{2}}\{2\}/\varepsilon_{\rm{2}}\{2\}$ are shown. The $p_{\rm{T}}$-integrated values for the constituent quark MC Glauber model chosen ($\rm{q}=3$) are observed to be similar in the left bottom panel of Fig.\  \ref{Fig4}. In that figure, the Xe--Xe centrality bin corresponds to the fourth point going left to right, while the Pb--Pb centrality bin corresponds to the third point. The ratio in the bottom panel  of Fig.\  \ref{Fig5} uses an interpolation of the Pb--Pb data points. The ratio is independent of the initial state model used, as all give very similar values of $\varepsilon_{\rm{2}}\{2\}$(Xe--Xe)/$\varepsilon_{\rm{2}}\{2\}$(Pb--Pb). Additionally, the transverse sizes ($R=\sqrt{S/\pi}$) are very similar, so the previously mentioned viscous corrections should cancel. The influence of radial flow should be very similar as $\langle p_{T}\rangle=0.710\pm0.004$ GeV/$c$ (Xe--Xe) and $\langle p_{T}\rangle=0.716\pm0.004$ GeV/$c$ (Pb--Pb) for charged hadrons \cite{Acharya:2018eaq}. The ratio is close to 1 and shows no significant $p_{\rm{T}}$ dependence. This indicates when such a scaling holds, it does so over the $p_{\rm{T}}$ range presented. This may show the $p_{\rm{T}}$-differential medium response ($\kappa_{2}(p_{\rm{T}})$) is controlled by the transverse density and size, independent of the collision system. A comparison of the scaled $p_{\rm{T}}$-differential coefficients for the same 30--40\% centrality bin from Xe--Xe and Pb--Pb collisions is also shown. In this case, the eccentricities are similar (the differences are within 1\%), however the transverse size and density of the Xe--Xe system is smaller. The ratio appears to mildly decrease with increasing $p_{\rm{T}}$. Whether this is the result of viscous effects related to the transverse size of the system influencing the mid-$p_{\rm{T}}$ region more, or a smaller radial flow in Xe--Xe, remains an open question.

\section{Summary}

The first measurements of anisotropic flow coefficients $v_{\rm{n}}$ in \sNNxs collisions from the ALICE detector at the LHC have been presented. Hydrodynamical predictions reproduce measurements of $v_{2}\{4\}$/$v_{2}\{2\}$ ratios from Xe--Xe collisions to within $\sim$ 15\% (Fig.\  \ref{Fig1}). In semi-central collisions, it is found that the $v_{2}\{2\}$ coefficient is lower in \sNNxs compared with \sNNps at the same centrality. The $v_3\{2\}$ coefficient is larger, consistent with expectations from hydrodynamical models that reproduce the differences for both systems within $\sim$ 5\% (Figs.\  \ref{Fig2} and \ref{Fig2e}). The differences for $v_2\{2\}$ are predicted to be driven largely by the hydrodynamical response of the system. For central collisions, $v_2\{2\}$ is found to be larger in Xe--Xe collisions, which agrees with predictions from hydrodynamic models, but the deviations tend to be larger than $\sim$ 5\% with respect to these models. The differences between two-particle $p_{\rm{T}}$-differential $v_{\rm{2}}\{2\}$ coefficients from Xe--Xe compared to Pb--Pb are found to be larger at mid-$p_{\rm{T}}$ compared to low-$p_{\rm{T}}$, whereas no such trend is observed for $v_{\rm{3}}\{2\}$ within uncertainties (Figs.\  \ref{Fig3v2} and \ref{Fig3v3}). The studies of the modeling of the initial state via eccentricity scaling with transverse density (Fig.\  \ref{Fig4}) have demonstrated that both the MC Glauber (constituent quarks) and the T$_{\rm{R}}$ENTo models provide the most satisfactory descriptions. However, the drop observed for $v_{\rm{2}}\{2\}/\varepsilon_{\rm{2}}\{2\}$ in central Xe--Xe and Pb--Pb collisions is not expected from hydrodynamic calculations. In the case of the MC Glauber implementations, the drop is more pronounced for nucleon and constituent quark ($\rm{q}=3$) sources, and may require some improvements in the initial state modeling for the region in Xe--Xe and Pb--Pb collisions where $\varepsilon_{\rm{2}}\{2\}$ has the largest contribution from initial state fluctuations. Finally, for two Xe--Xe and Pb--Pb centrality bins with a similar transverse density and size, it is found that the double ratio [$v_{\rm{2}}\{2\}/\varepsilon_{\rm{2}}\{2\}$(Xe--Xe)]/[$v_{\rm{2}}\{2\}/\varepsilon_{\rm{2}}\{2\}$(Pb--Pb)] is largely independent of $p_{\rm{T}}$ (Fig.\  \ref{Fig5}). This may indicate the $p_{\rm{T}}$-differential medium response is controlled by the transverse density and size, independent of the collision system.

\newenvironment{acknowledgement}{\relax}{\relax}
\begin{acknowledgement}
\section*{Acknowledgements}

The ALICE Collaboration would like to thank all its engineers and technicians for their invaluable contributions to the construction of the experiment and the CERN accelerator teams for the outstanding performance of the LHC complex.
The ALICE Collaboration gratefully acknowledges the resources and support provided by all Grid centres and the Worldwide LHC Computing Grid (WLCG) collaboration.
The ALICE Collaboration acknowledges the following funding agencies for their support in building and running the ALICE detector:
A. I. Alikhanyan National Science Laboratory (Yerevan Physics Institute) Foundation (ANSL), State Committee of Science and World Federation of Scientists (WFS), Armenia;
Austrian Academy of Sciences and Nationalstiftung f\"{u}r Forschung, Technologie und Entwicklung, Austria;
Ministry of Communications and High Technologies, National Nuclear Research Center, Azerbaijan;
Conselho Nacional de Desenvolvimento Cient\'{\i}fico e Tecnol\'{o}gico (CNPq), Universidade Federal do Rio Grande do Sul (UFRGS), Financiadora de Estudos e Projetos (Finep) and Funda\c{c}\~{a}o de Amparo \`{a} Pesquisa do Estado de S\~{a}o Paulo (FAPESP), Brazil;
Ministry of Science \& Technology of China (MSTC), National Natural Science Foundation of China (NSFC) and Ministry of Education of China (MOEC) , China;
Ministry of Science and Education, Croatia;
Ministry of Education, Youth and Sports of the Czech Republic, Czech Republic;
The Danish Council for Independent Research | Natural Sciences, the Carlsberg Foundation and Danish National Research Foundation (DNRF), Denmark;
Helsinki Institute of Physics (HIP), Finland;
Commissariat \`{a} l'Energie Atomique (CEA) and Institut National de Physique Nucl\'{e}aire et de Physique des Particules (IN2P3) and Centre National de la Recherche Scientifique (CNRS), France;
Bundesministerium f\"{u}r Bildung, Wissenschaft, Forschung und Technologie (BMBF) and GSI Helmholtzzentrum f\"{u}r Schwerionenforschung GmbH, Germany;
General Secretariat for Research and Technology, Ministry of Education, Research and Religions, Greece;
National Research, Development and Innovation Office, Hungary;
Department of Atomic Energy Government of India (DAE), Department of Science and Technology, Government of India (DST), University Grants Commission, Government of India (UGC) and Council of Scientific and Industrial Research (CSIR), India;
Indonesian Institute of Science, Indonesia;
Centro Fermi - Museo Storico della Fisica e Centro Studi e Ricerche Enrico Fermi and Istituto Nazionale di Fisica Nucleare (INFN), Italy;
Institute for Innovative Science and Technology , Nagasaki Institute of Applied Science (IIST), Japan Society for the Promotion of Science (JSPS) KAKENHI and Japanese Ministry of Education, Culture, Sports, Science and Technology (MEXT), Japan;
Consejo Nacional de Ciencia (CONACYT) y Tecnolog\'{i}a, through Fondo de Cooperaci\'{o}n Internacional en Ciencia y Tecnolog\'{i}a (FONCICYT) and Direcci\'{o}n General de Asuntos del Personal Academico (DGAPA), Mexico;
Nederlandse Organisatie voor Wetenschappelijk Onderzoek (NWO), Netherlands;
The Research Council of Norway, Norway;
Commission on Science and Technology for Sustainable Development in the South (COMSATS), Pakistan;
Pontificia Universidad Cat\'{o}lica del Per\'{u}, Peru;
Ministry of Science and Higher Education and National Science Centre, Poland;
Korea Institute of Science and Technology Information and National Research Foundation of Korea (NRF), Republic of Korea;
Ministry of Education and Scientific Research, Institute of Atomic Physics and Romanian National Agency for Science, Technology and Innovation, Romania;
Joint Institute for Nuclear Research (JINR), Ministry of Education and Science of the Russian Federation and National Research Centre Kurchatov Institute, Russia;
Ministry of Education, Science, Research and Sport of the Slovak Republic, Slovakia;
National Research Foundation of South Africa, South Africa;
Centro de Aplicaciones Tecnol\'{o}gicas y Desarrollo Nuclear (CEADEN), Cubaenerg\'{\i}a, Cuba and Centro de Investigaciones Energ\'{e}ticas, Medioambientales y Tecnol\'{o}gicas (CIEMAT), Spain;
Swedish Research Council (VR) and Knut \& Alice Wallenberg Foundation (KAW), Sweden;
European Organization for Nuclear Research, Switzerland;
National Science and Technology Development Agency (NSDTA), Suranaree University of Technology (SUT) and Office of the Higher Education Commission under NRU project of Thailand, Thailand;
Turkish Atomic Energy Agency (TAEK), Turkey;
National Academy of  Sciences of Ukraine, Ukraine;
Science and Technology Facilities Council (STFC), United Kingdom;
National Science Foundation of the United States of America (NSF) and United States Department of Energy, Office of Nuclear Physics (DOE NP), United States of America.    
\end{acknowledgement}

\bibliographystyle{utphys}   
\bibliography{cumuRefs}

\newpage
\appendix
\section{The ALICE Collaboration}
\label{app:collab}

\begingroup
\small
\begin{flushleft}
S.~Acharya\Irefn{org138}\And 
F.T.-.~Acosta\Irefn{org22}\And 
D.~Adamov\'{a}\Irefn{org94}\And 
J.~Adolfsson\Irefn{org81}\And 
M.M.~Aggarwal\Irefn{org98}\And 
G.~Aglieri Rinella\Irefn{org36}\And 
M.~Agnello\Irefn{org33}\And 
N.~Agrawal\Irefn{org49}\And 
Z.~Ahammed\Irefn{org138}\And 
S.U.~Ahn\Irefn{org77}\And 
S.~Aiola\Irefn{org143}
% ...
\And
\renewcommand\labelenumi{\textsuperscript{\theenumi}~}

\section*{Affiliation notes}
\renewcommand\theenumi{\roman{enumi}}
\begin{Authlist}
\item \Adef{org*}Deceased
\item \Adef{orgI}Dipartimento DET del Politecnico di Torino, Turin, Italy
\item \Adef{orgII}M.V. Lomonosov Moscow State University, D.V. Skobeltsyn Institute of Nuclear, Physics, Moscow, Russia
\item \Adef{orgIII}Department of Applied Physics, Aligarh Muslim University, Aligarh, India
\item \Adef{orgIV}Institute of Theoretical Physics, University of Wroclaw, Poland
\end{Authlist}

\section*{Collaboration Institutes}
\renewcommand\theenumi{\arabic{enumi}~}
\begin{Authlist}
\item \Idef{org1}A.I. Alikhanyan National Science Laboratory (Yerevan Physics Institute) Foundation, Yerevan, Armenia
\item \Idef{org2}Benem\'{e}rita Universidad Aut\'{o}noma de Puebla, Puebla, Mexico
\item \Idef{org3}Bogolyubov Institute for Theoretical Physics, National Academy of Sciences of Ukraine, Kiev, Ukraine
\item \Idef{org4}Bose Institute, Department of Physics  and Centre for Astroparticle Physics and Space Science (CAPSS), Kolkata, India
\item \Idef{org5}Budker Institute for Nuclear Physics, Novosibirsk, Russia
\item \Idef{org6}California Polytechnic State University, San Luis Obispo, California, United States
\item \Idef{org7}Central China Normal University, Wuhan, China
\item \Idef{org8}Centre de Calcul de l'IN2P3, Villeurbanne, Lyon, France
\item \Idef{org9}Centro de Aplicaciones Tecnol\'{o}gicas y Desarrollo Nuclear (CEADEN), Havana, Cuba
\item \Idef{org10}Centro de Investigaci\'{o}n y de Estudios Avanzados (CINVESTAV), Mexico City and M\'{e}rida, Mexico
\item \Idef{org11}Centro Fermi - Museo Storico della Fisica e Centro Studi e Ricerche ``Enrico Fermi', Rome, Italy
\item \Idef{org12}Chicago State University, Chicago, Illinois, United States
\item \Idef{org13}China Institute of Atomic Energy, Beijing, China
\item \Idef{org14}Chonbuk National University, Jeonju, Republic of Korea
\item \Idef{org15}Comenius University Bratislava, Faculty of Mathematics, Physics and Informatics, Bratislava, Slovakia
\item \Idef{org16}COMSATS Institute of Information Technology (CIIT), Islamabad, Pakistan
\item \Idef{org17}Creighton University, Omaha, Nebraska, United States
\item \Idef{org18}Department of Physics, Aligarh Muslim University, Aligarh, India
\item \Idef{org19}Department of Physics, Ohio State University, Columbus, Ohio, United States
\item \Idef{org20}Department of Physics, Pusan National University, Pusan, Republic of Korea
\item \Idef{org21}Department of Physics, Sejong University, Seoul, Republic of Korea
\item \Idef{org22}Department of Physics, University of California, Berkeley, California, United States
\item \Idef{org23}Department of Physics, University of Oslo, Oslo, Norway
\item \Idef{org24}Department of Physics and Technology, University of Bergen, Bergen, Norway
\item \Idef{org25}Dipartimento di Fisica dell'Universit\`{a} 'La Sapienza' and Sezione INFN, Rome, Italy
\item \Idef{org26}Dipartimento di Fisica dell'Universit\`{a} and Sezione INFN, Cagliari, Italy
\item \Idef{org27}Dipartimento di Fisica dell'Universit\`{a} and Sezione INFN, Trieste, Italy
\item \Idef{org28}Dipartimento di Fisica dell'Universit\`{a} and Sezione INFN, Turin, Italy
\item \Idef{org29}Dipartimento di Fisica e Astronomia dell'Universit\`{a} and Sezione INFN, Bologna, Italy
\item \Idef{org30}Dipartimento di Fisica e Astronomia dell'Universit\`{a} and Sezione INFN, Catania, Italy
\item \Idef{org31}Dipartimento di Fisica e Astronomia dell'Universit\`{a} and Sezione INFN, Padova, Italy
\item \Idef{org32}Dipartimento di Fisica `E.R.~Caianiello' dell'Universit\`{a} and Gruppo Collegato INFN, Salerno, Italy
\item \Idef{org33}Dipartimento DISAT del Politecnico and Sezione INFN, Turin, Italy
\item \Idef{org34}Dipartimento di Scienze e Innovazione Tecnologica dell'Universit\`{a} del Piemonte Orientale and INFN Sezione di Torino, Alessandria, Italy
\item \Idef{org35}Dipartimento Interateneo di Fisica `M.~Merlin' and Sezione INFN, Bari, Italy
\item \Idef{org36}European Organization for Nuclear Research (CERN), Geneva, Switzerland
\item \Idef{org37}Faculty of Electrical Engineering, Mechanical Engineering and Naval Architecture, University of Split, Split, Croatia
\item \Idef{org38}Faculty of Engineering and Science, Western Norway University of Applied Sciences, Bergen, Norway
\item \Idef{org39}Faculty of Nuclear Sciences and Physical Engineering, Czech Technical University in Prague, Prague, Czech Republic
\item \Idef{org40}Faculty of Science, P.J.~\v{S}af\'{a}rik University, Ko\v{s}ice, Slovakia
\item \Idef{org41}Frankfurt Institute for Advanced Studies, Johann Wolfgang Goethe-Universit\"{a}t Frankfurt, Frankfurt, Germany
\item \Idef{org42}Gangneung-Wonju National University, Gangneung, Republic of Korea
\item \Idef{org43}Gauhati University, Department of Physics, Guwahati, India
\item \Idef{org44}Helmholtz-Institut f\"{u}r Strahlen- und Kernphysik, Rheinische Friedrich-Wilhelms-Universit\"{a}t Bonn, Bonn, Germany
\item \Idef{org45}Helsinki Institute of Physics (HIP), Helsinki, Finland
\item \Idef{org46}Hiroshima University, Hiroshima, Japan
\item \Idef{org47}Hochschule Worms, Zentrum  f\"{u}r Technologietransfer und Telekommunikation (ZTT), Worms, Germany
\item \Idef{org48}Horia Hulubei National Institute of Physics and Nuclear Engineering, Bucharest, Romania
\item \Idef{org49}Indian Institute of Technology Bombay (IIT), Mumbai, India
\item \Idef{org50}Indian Institute of Technology Indore, Indore, India
\item \Idef{org51}Indonesian Institute of Sciences, Jakarta, Indonesia
\item \Idef{org52}INFN, Laboratori Nazionali di Frascati, Frascati, Italy
\item \Idef{org53}INFN, Sezione di Bari, Bari, Italy
\item \Idef{org54}INFN, Sezione di Bologna, Bologna, Italy
\item \Idef{org55}INFN, Sezione di Cagliari, Cagliari, Italy
\item \Idef{org56}INFN, Sezione di Catania, Catania, Italy
\item \Idef{org57}INFN, Sezione di Padova, Padova, Italy
\item \Idef{org58}INFN, Sezione di Roma, Rome, Italy
\item \Idef{org59}INFN, Sezione di Torino, Turin, Italy
\item \Idef{org60}INFN, Sezione di Trieste, Trieste, Italy
\item \Idef{org61}Inha University, Incheon, Republic of Korea
\item \Idef{org62}Institut de Physique Nucl\'{e}aire d'Orsay (IPNO), Institut National de Physique Nucl\'{e}aire et de Physique des Particules (IN2P3/CNRS), Universit\'{e} de Paris-Sud, Universit\'{e} Paris-Saclay, Orsay, France
\item \Idef{org63}Institute for Nuclear Research, Academy of Sciences, Moscow, Russia
\item \Idef{org64}Institute for Subatomic Physics, Utrecht University/Nikhef, Utrecht, Netherlands
\item \Idef{org65}Institute for Theoretical and Experimental Physics, Moscow, Russia
\item \Idef{org66}Institute of Experimental Physics, Slovak Academy of Sciences, Ko\v{s}ice, Slovakia
\item \Idef{org67}Institute of Physics, Bhubaneswar, India
\item \Idef{org68}Institute of Physics of the Czech Academy of Sciences, Prague, Czech Republic
\item \Idef{org69}Institute of Space Science (ISS), Bucharest, Romania
\item \Idef{org70}Institut f\"{u}r Kernphysik, Johann Wolfgang Goethe-Universit\"{a}t Frankfurt, Frankfurt, Germany
\item \Idef{org71}Instituto de Ciencias Nucleares, Universidad Nacional Aut\'{o}noma de M\'{e}xico, Mexico City, Mexico
\item \Idef{org72}Instituto de F\'{i}sica, Universidade Federal do Rio Grande do Sul (UFRGS), Porto Alegre, Brazil
\item \Idef{org73}Instituto de F\'{\i}sica, Universidad Nacional Aut\'{o}noma de M\'{e}xico, Mexico City, Mexico
\item \Idef{org74}iThemba LABS, National Research Foundation, Somerset West, South Africa
\item \Idef{org75}Johann-Wolfgang-Goethe Universit\"{a}t Frankfurt Institut f\"{u}r Informatik, Fachbereich Informatik und Mathematik, Frankfurt, Germany
\item \Idef{org76}Joint Institute for Nuclear Research (JINR), Dubna, Russia
\item \Idef{org77}Korea Institute of Science and Technology Information, Daejeon, Republic of Korea
\item \Idef{org78}KTO Karatay University, Konya, Turkey
\item \Idef{org79}Laboratoire de Physique Subatomique et de Cosmologie, Universit\'{e} Grenoble-Alpes, CNRS-IN2P3, Grenoble, France
\item \Idef{org80}Lawrence Berkeley National Laboratory, Berkeley, California, United States
\item \Idef{org81}Lund University Department of Physics, Division of Particle Physics, Lund, Sweden
\item \Idef{org82}Nagasaki Institute of Applied Science, Nagasaki, Japan
\item \Idef{org83}Nara Women{'}s University (NWU), Nara, Japan
\item \Idef{org84}National and Kapodistrian University of Athens, School of Science, Department of Physics , Athens, Greece
\item \Idef{org85}National Centre for Nuclear Research, Warsaw, Poland
\item \Idef{org86}National Institute of Science Education and Research, HBNI, Jatni, India
\item \Idef{org87}National Nuclear Research Center, Baku, Azerbaijan
\item \Idef{org88}National Research Centre Kurchatov Institute, Moscow, Russia
\item \Idef{org89}Niels Bohr Institute, University of Copenhagen, Copenhagen, Denmark
\item \Idef{org90}Nikhef, National institute for subatomic physics, Amsterdam, Netherlands
\item \Idef{org91}NRC Kurchatov Institute IHEP, Protvino, Russia
\item \Idef{org92}NRNU Moscow Engineering Physics Institute, Moscow, Russia
\item \Idef{org93}Nuclear Physics Group, STFC Daresbury Laboratory, Daresbury, United Kingdom
\item \Idef{org94}Nuclear Physics Institute of the Czech Academy of Sciences, \v{R}e\v{z} u Prahy, Czech Republic
\item \Idef{org95}Oak Ridge National Laboratory, Oak Ridge, Tennessee, United States
\item \Idef{org96}Petersburg Nuclear Physics Institute, Gatchina, Russia
\item \Idef{org97}Physics department, Faculty of science, University of Zagreb, Zagreb, Croatia
\item \Idef{org98}Physics Department, Panjab University, Chandigarh, India
\item \Idef{org99}Physics Department, University of Jammu, Jammu, India
\item \Idef{org100}Physics Department, University of Rajasthan, Jaipur, India
\item \Idef{org101}Physikalisches Institut, Eberhard-Karls-Universit\"{a}t T\"{u}bingen, T\"{u}bingen, Germany
\item \Idef{org102}Physikalisches Institut, Ruprecht-Karls-Universit\"{a}t Heidelberg, Heidelberg, Germany
\item \Idef{org103}Physik Department, Technische Universit\"{a}t M\"{u}nchen, Munich, Germany
\item \Idef{org104}Research Division and ExtreMe Matter Institute EMMI, GSI Helmholtzzentrum f\"ur Schwerionenforschung GmbH, Darmstadt, Germany
\item \Idef{org105}Rudjer Bo\v{s}kovi\'{c} Institute, Zagreb, Croatia
\item \Idef{org106}Russian Federal Nuclear Center (VNIIEF), Sarov, Russia
\item \Idef{org107}Saha Institute of Nuclear Physics, Kolkata, India
\item \Idef{org108}School of Physics and Astronomy, University of Birmingham, Birmingham, United Kingdom
\item \Idef{org109}Secci\'{o}n F\'{\i}sica, Departamento de Ciencias, Pontificia Universidad Cat\'{o}lica del Per\'{u}, Lima, Peru
\item \Idef{org110}Shanghai Institute of Applied Physics, Shanghai, China
\item \Idef{org111}Stefan Meyer Institut f\"{u}r Subatomare Physik (SMI), Vienna, Austria
\item \Idef{org112}SUBATECH, IMT Atlantique, Universit\'{e} de Nantes, CNRS-IN2P3, Nantes, France
\item \Idef{org113}Suranaree University of Technology, Nakhon Ratchasima, Thailand
\item \Idef{org114}Technical University of Ko\v{s}ice, Ko\v{s}ice, Slovakia
\item \Idef{org115}Technische Universit\"{a}t M\"{u}nchen, Excellence Cluster 'Universe', Munich, Germany
\item \Idef{org116}The Henryk Niewodniczanski Institute of Nuclear Physics, Polish Academy of Sciences, Cracow, Poland
\item \Idef{org117}The University of Texas at Austin, Austin, Texas, United States
\item \Idef{org118}Universidad Aut\'{o}noma de Sinaloa, Culiac\'{a}n, Mexico
\item \Idef{org119}Universidade de S\~{a}o Paulo (USP), S\~{a}o Paulo, Brazil
\item \Idef{org120}Universidade Estadual de Campinas (UNICAMP), Campinas, Brazil
\item \Idef{org121}Universidade Federal do ABC, Santo Andre, Brazil
\item \Idef{org122}University College of Southeast Norway, Tonsberg, Norway
\item \Idef{org123}University of Cape Town, Cape Town, South Africa
\item \Idef{org124}University of Houston, Houston, Texas, United States
\item \Idef{org125}University of Jyv\"{a}skyl\"{a}, Jyv\"{a}skyl\"{a}, Finland
\item \Idef{org126}University of Liverpool, Department of Physics Oliver Lodge Laboratory , Liverpool, United Kingdom
\item \Idef{org127}University of Tennessee, Knoxville, Tennessee, United States
\item \Idef{org128}University of the Witwatersrand, Johannesburg, South Africa
\item \Idef{org129}University of Tokyo, Tokyo, Japan
\item \Idef{org130}University of Tsukuba, Tsukuba, Japan
\item \Idef{org131}Universit\'{e} Clermont Auvergne, CNRS/IN2P3, LPC, Clermont-Ferrand, France
\item \Idef{org132}Universit\'{e} de Lyon, Universit\'{e} Lyon 1, CNRS/IN2P3, IPN-Lyon, Villeurbanne, Lyon, France
\item \Idef{org133}Universit\'{e} de Strasbourg, CNRS, IPHC UMR 7178, F-67000 Strasbourg, France, Strasbourg, France
\item \Idef{org134} Universit\'{e} Paris-Saclay Centre d¿\'Etudes de Saclay (CEA), IRFU, Department de Physique Nucl\'{e}aire (DPhN), Saclay, France
\item \Idef{org135}Universit\`{a} degli Studi di Pavia, Pavia, Italy
\item \Idef{org136}Universit\`{a} di Brescia, Brescia, Italy
\item \Idef{org137}V.~Fock Institute for Physics, St. Petersburg State University, St. Petersburg, Russia
\item \Idef{org138}Variable Energy Cyclotron Centre, Kolkata, India
\item \Idef{org139}Warsaw University of Technology, Warsaw, Poland
\item \Idef{org140}Wayne State University, Detroit, Michigan, United States
\item \Idef{org141}Westf\"{a}lische Wilhelms-Universit\"{a}t M\"{u}nster, Institut f\"{u}r Kernphysik, M\"{u}nster, Germany
\item \Idef{org142}Wigner Research Centre for Physics, Hungarian Academy of Sciences, Budapest, Hungary
\item \Idef{org143}Yale University, New Haven, Connecticut, United States
\item \Idef{org144}Yonsei University, Seoul, Republic of Korea
\end{Authlist}
\endgroup
\end{document}